\documentclass[a4paper]{article}
\usepackage[inline]{enumitem}
\usepackage{multirow}
\usepackage{multicol}

\usepackage[utf8]{inputenc}

\usepackage{INTERSPEECH2020}
\usepackage[font=small,skip=1pt]{caption}

\usepackage{cleveref}

\title{Voice activity detection in the wild via weakly supervised sound event detection}
\name{Yefei Chen\textsuperscript{*}, Heinrich Dinkel\textsuperscript{*}, Mengyue Wu ,Kai Yu\thanks{\textsuperscript{*} equal contribution. Mengyue Wu and Kai Yu are the corresponding authors.}}
\address{
MoE Key Lab of Artificial Intelligence\\
SpeechLab, Department of Computer Science and Engineering\\
AI Institute, Shanghai Jiao Tong University, Shanghai, China
  }
\email{\{chenyefei,richman,mengyuewu, kai.yu\}@sjtu.edu.cn}

\begin{document}

\maketitle
\begin{abstract}
Traditional supervised voice activity detection (VAD) methods work well in clean and controlled scenarios, with performance severely degrading in real-world applications. 
One possible bottleneck is that speech in the wild contains unpredictable noise types, hence frame-level label prediction is difficult, which is required for traditional supervised VAD training. 
In contrast, we propose a general-purpose VAD (GPVAD) framework, which can be easily trained from noisy data in a weakly supervised fashion, requiring only clip-level labels. 
We proposed two GPVAD models, one full (GPV-F), trained on 527 Audioset sound events, and one binary (GPV-B), only distinguishing speech and noise.  
We evaluate the two GPV models against a CRNN based standard VAD model (VAD-C) on three different evaluation protocols (clean, synthetic noise, real data).
Results show that our proposed GPV-F demonstrates competitive performance in clean and synthetic scenarios compared to traditional VAD-C.
Further, in real-world evaluation, GPV-F largely outperforms VAD-C in terms of frame-level evaluation metrics as well as segment-level ones.
With a much lower requirement for frame-labeled data, the naive binary clip-level GPV-B model can still achieve comparable performance to VAD-C in real-world scenarios.
\end{abstract}

\section{Introduction}

Voice activity detection (VAD), whose main objective is to detect voiced speech segments and distinguish those from unvoiced ones, is a crucial component for tasks such as speech recognition, speaker recognition, and speaker verification.
Deep learning approaches have been successfully applied to VAD~\cite{Hughes2013,ryant2013speech,thomas2014analyzing,Lavechin2020}. 
For VAD in complex environments, neural networks (NN) have been successful.
Deep neural networks (DNN) and specifically convolutional neural networks (CNN) offer improved modeling capabilities compared to traditional methods~\cite{ryant2013speech}, while recurrent- (RNN) and long short-term memory (LSTM) networks can better model long-term dependencies between sequential inputs~\cite{Hughes2013, eyben2013real,Tong2016}.
However, despite the application of deep learning methods, NN-based VAD training still requires frame labels. 
Thus training data utilized is usually under controlled environment with or without additional synthetic noise~\cite{hirsch2000aurora}. This inevitably prevents VAD from real-world applications, where speech in the wild is often accompanied by countless unseen noises with different features. 

\begin{figure*}
    \centering
    \includegraphics[width=\textwidth]{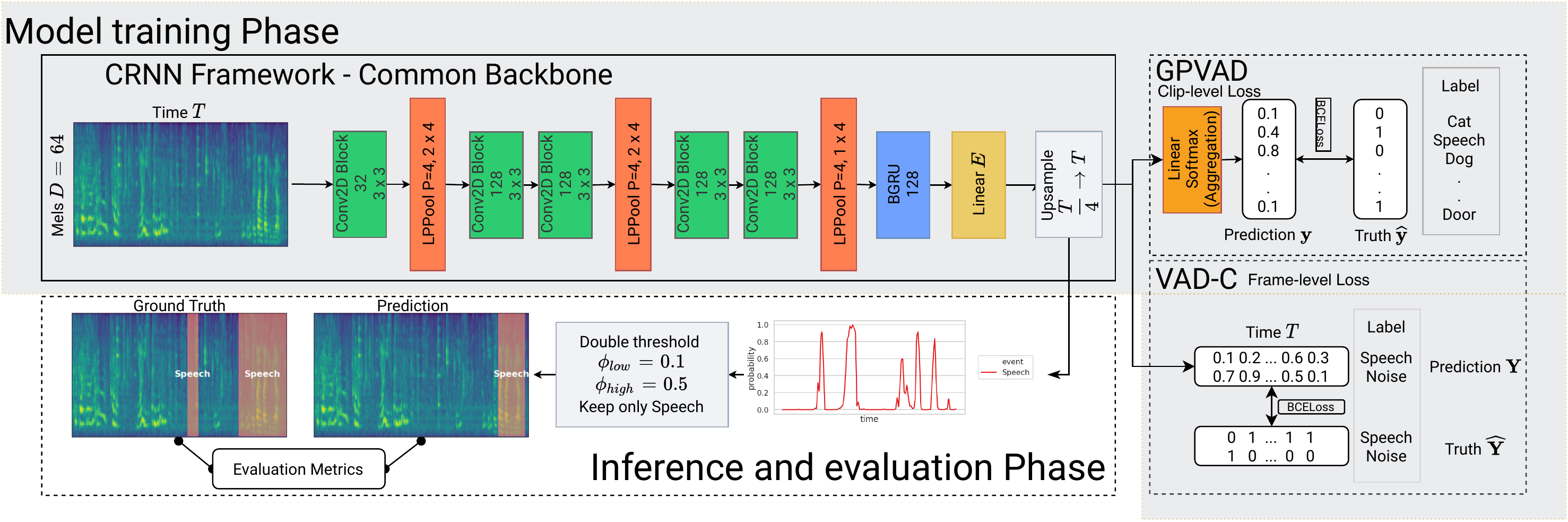}
    \caption{The proposed framework. A CRNN architecture is utilized, while GPVAD is trained clip-level labels, and VAD-C trained on frame-level labels. Each Conv2d block represents a batch-normalization, followed by a zero-padded 2-dimensional convolution with kernel size $3\times3$ and a leaky ReLU activation with a negative slope of $0.1$.
    The CNN output is fed into a bidirectional gated recurrent unit (GRU) with 128 hidden units.
    The architecture sub-samples the temporal dimension $T$ by a factor of 4 and later upsampled to match the original input temporal dimension.
    The number of events $E$ is set to be $527$ for GPV-F, $2$ for GPV-B, and VAD-C.
    After post-processing the output, only the \textit{Speech} event is kept for final evaluation.}
    \label{fig:model_approach}
\end{figure*}

Therefore, this paper intends to propose a method to detect speech beyond clean and controlled noisy environment. It should be noted that frame-level labels are quite unlikely to come with real-world recordings since manual labeling is costly, and label predictions from a Hidden Markov model need prior knowledge about the language being spoken~\cite{rabiner_hmm_tutorial}. 
A task to detect speech components while enabling noisy data training, is related to weakly-supervised sound event detection (WSSED), which detects and localizes different sounds, including speech via clip-level supervision. 
Since WSSED systems are reported~\cite{dinkel2019duration} to be robust to noise and only require clip-level labels, this work integrates WSSED methods in scaling VAD to speech in-the-wild scenarios and relaxing its dependence on frame labeling. Specifically, we investigate two questions: 1) Are current, multi-class WSSED models comparable in performance to DNN-based VAD; 2) Is utterance-level training a viable alternative compared to frame-level?
We thus introduce our framework, a general-purpose training framework for VAD (GPVAD, see \Cref{fig:model_approach}).
By general purpose, we refer to two distinct aspects: First, the framework is noise-robust and capable of being deployed in wild, real-world scenarios; Secondly, the framework can be trained on unconstrained data, thus enabling learning from big webly data like noisy online-videos.

The paper is structured as follows: In \Cref{sec:relatedwork}, we briefly review the related work on WSSED and how it can be transferred for VAD in the wild.
In \Cref{sec:approch}, the GPVAD approach is introduced. 
Moreover, in \Cref{sec:experiments} we introduce our experimental setup and provide implementation details. 
In \Cref{sec:results} the results are presented and finally in \Cref{sec:conclusion} a conclusion is provided.

\section{Weakly supervised sound event detection}
\label{sec:relatedwork}
Since WSSED can work well in detecting speech in a noisy environment without frame-level labeling, we borrow this idea to realize VAD in the wild. Here we present related work on sound event detection (SED), which aims to classify (audio tagging) and possibly localize multiple co-occurring sound events from a given audio clip.
In this work, we mainly focus on weakly-supervised SED (WSSED), a semi-supervised task, which has only access to clip-level labels during training, yet needs to classify and localize a specific event during evaluation. This weakly-supervised fashion enables training on noisy data with lower requirements for the labeling method.
Recent advances in weakly supervised sound event detection, in particular, the detection and classification of acoustic scenes and events (DCASE) challenges~\cite{Serizel2018}, led to large improvements for predicting accurate sound event boundaries as well as event-labels~\cite{lin2019specialized,yong_xu_att,Cakr2017,Kong2018,Pellegrini2019,Kong2019}.
In particular, recent work~\cite{dinkel2019duration} has shown promising performance regarding short, sporadic events such as speech.

\section{VAD in the wild via WSSED}
\label{sec:approch}

Traditionally, VAD for noisy scenarios is modeled as in \Cref{eq:noisy_eq}.
The assumption is that additive noise $\mathbf{u}$ can be filtered out from an observed speech signal $\mathbf{x}$ to obtain clean speech $\mathbf{s}$.
\begin{equation}
\label{eq:noisy_eq}
    \mathbf{x} = \mathbf{s} + \mathbf{u}
\end{equation}
However, directly modeling $\mathbf{u}$ is rather tricky, since each type of noise has its individual traits.
Therefore, we aim at learning the properties of $\mathbf{s}$ by observing it with potentially $L$ different non-speech events $\left( \mathbf{u}_1 \ldots, \mathbf{u}_L \right)$.
Those events are not restricted to being background/foreground noises and can have distinct real-world sounds (e.g., Cat, Music).

\begin{align}
\label{eq:model}
    \begin{split}
        \mathcal{X} &= \{ \mathbf{x}_1,\ldots, \mathbf{x}_{l}, \ldots , \mathbf{x}_{L} \}\\
        \mathbf{x}_l &= \left( \mathbf{s}, \mathbf{u}_l \right)
    \end{split}
\end{align}

Our approach stems from multiple instance learning (MIL), meaning that training set knowledge about specific labels is incomplete (e.g., Speech never directly observed).
Here, we model our observed speech data $\mathcal{X}$ as a ``bag'', containing all co-occurrences of Speech in conjunction with another, possibly noisy background/foreground event label $l \in \{1, \ldots, L\}$ from a set of all possible event labels $L < E$ (\Cref{eq:model}).
So to speak, our approach aims to refine a model's belief about the speech signal $\mathbf{s}$, within complex environmental scenarios.
The advantage of this modeling method is that it can be applied for both frame- and clip-level training.
Our GPVAD, therefore, relaxes these constraints by allowing training on clip/utterance level, where each training clip contains at least one event of interest.
We propose two different models: GPV-F, which outputs $E=527$ labels ($L=405$) and the naive GPV-B, $E=2, L=1$.
GPV-F can be seen as a full-fledged WSSED approach using maximal label supervision and is, therefore, more costly than GPV-B, which only requires knowledge about a clip containing Speech.
However, GPV-F should be capable of modeling each individual noise-event instead of clustering all noise into a single class (GPV-B), thus possibly enhancing performance in heavy noise scenarios.
The two models are compared against a model trained on frame-level, further referred to as VAD-C.

All models share a common backbone convolutional recurrent neural network (CRNN)~\cite{dinkel2019duration} approach used in WSSED, which is shown to be robust towards short, sporadic events such as Speech.
The following modification to~\cite{dinkel2019duration} have been done: \begin{enumerate*}
    \item Add an upsampling operation, such that the models' time-resolution remains constant.
    \item Use $L^p$ pooling as our default with $p=4$, as it has been seen to be beneficial for duration invariant estimates.
\end{enumerate*}
Different from VAD-C training, where frame-level labels are available, our GPVAD framework is split into two distinct stages.
During training, only clip/utterance-level labels are accessible. 
Therefore a temporal pooling function is required (\Cref{eq:linear_softmax}).
During inference, post-processing needs to be applied (\Cref{para:post_processing}) to convert probability sequences into binary labels (absence/presence of an event) as well as any predicted non-speech label is discarded.
The framework is depicted in \Cref{fig:model_approach}.



\section{Experiments}
\label{sec:experiments}

%

In our work, deep neural networks were implemented in PyTorch ~\cite{paszke2017automatic}, front-end feature extraction utilized librosa~\cite{Librosa_mcfee}. 
Code is available online\footnote{Available at github.com/richermans/gpv}.

\subsection{Datasets}
\label{ssec:data}




\begin{table}[htbp]
    \centering
    \begin{tabular}{l|r|r|r|r}
        Datatype & Name   & Condition & Label & Duration  \\
        \hline\hline
        \multirow{2}{*}{Training} & Audioset & Real & Clip &  15 h \\
        & Aurora 4+ & Syn & Frame &  30 h \\
        \hline
        \multirow{3}{*}{Evaluation}& Aurora 4 (A) & Clean & Frame & 40 min \\
        & Aurora 4 (B) & Syn & Frame & 8.7 h \\
        & DCASE18 &  Real & Frame &  100 min \\
    \end{tabular}
    \caption{Training datasets for GPVAD (Audioset) and VAD-C (Aurora 4+), as well as the three proposed evaluation protocols for clean, synthetic noise and real-world scenarios. Duration represents the approximate length of speech.}
    \label{tab:datasets}
\end{table}

Utilized datasets in this work can be split into a train data portion, which differs between the GPVAD and VAD approaches, and evaluation data, which is shared by both approaches.
Our main GPVAD training dataset is the ``balanced" set provided by the AudioSet corpus~\cite{Gemmeke2017}, containing 21100/22160 (due to unavailability) 10-second Youtube audio clips, categorized into 527 noisy event labels.
From the available 21100 clips (58h), 5452 clips ($\approx$ 15h) are labeled as containing speech, but always alongside $L=405$ other events (e.g., Bark).
Regarding GPV-B, we replace all 526 events in the balanced dataset, not being speech as ``noise'', thus $\mathcal{X}_{\text{GPV-B}} = \{ \left(\mathbf{s}, \mathbf{u}_{noise}, \right), \mathbf{u}_{noise} \}$. 
It is important to note that for GPV-B/V training, speech is never individually observed.

Our VAD-C model is trained on the Aurora 4 training set extended by 15 hours of Switchboard~\cite{godfrey_switchboard}, obtaining our Aurora 4+ training subset, containing clean as well as synthetic noise data.
The additive synthetic noise (Syn) is obtained from six different noise types (car, babble, restaurant, street, airport, and train) that were added at randomly selected SNRs between 10 and 20 dB.
All utilized datasets are described in \Cref{tab:datasets}.
Three different evaluation scenarios are proposed. 
First, we validate on the 40 minutes long, clean Aurora 4 test set~\cite{hirsch2000aurora}.
Second, we synthesize a noisy test set based on the clean Aurora 4 test set by randomly adding noise from 100 noise types using a SNR ranging from 5db to 15db in steps of 1db.
Lastly, we merge the development and evaluation tracks of the DCASE18 challenge~\cite{Serizel2018}, itself a subset of Audioset, to create our real-world evaluation data.
The DCASE18 data provides ten domestic environment event labels, of which we neglect all labels other than Speech, but report the number of instances where non-speech labels were present.
Our DCASE18 evaluation set encompasses 596 utterances labelled as "Speech", 414 utterances (69\%) contain another non-speech label, 114 utterances (20\%) only contain speech and 68 utterances (11\%) contain two or more non-speech labels.

\begin{figure}
    \centering
    \includegraphics[width=\linewidth]{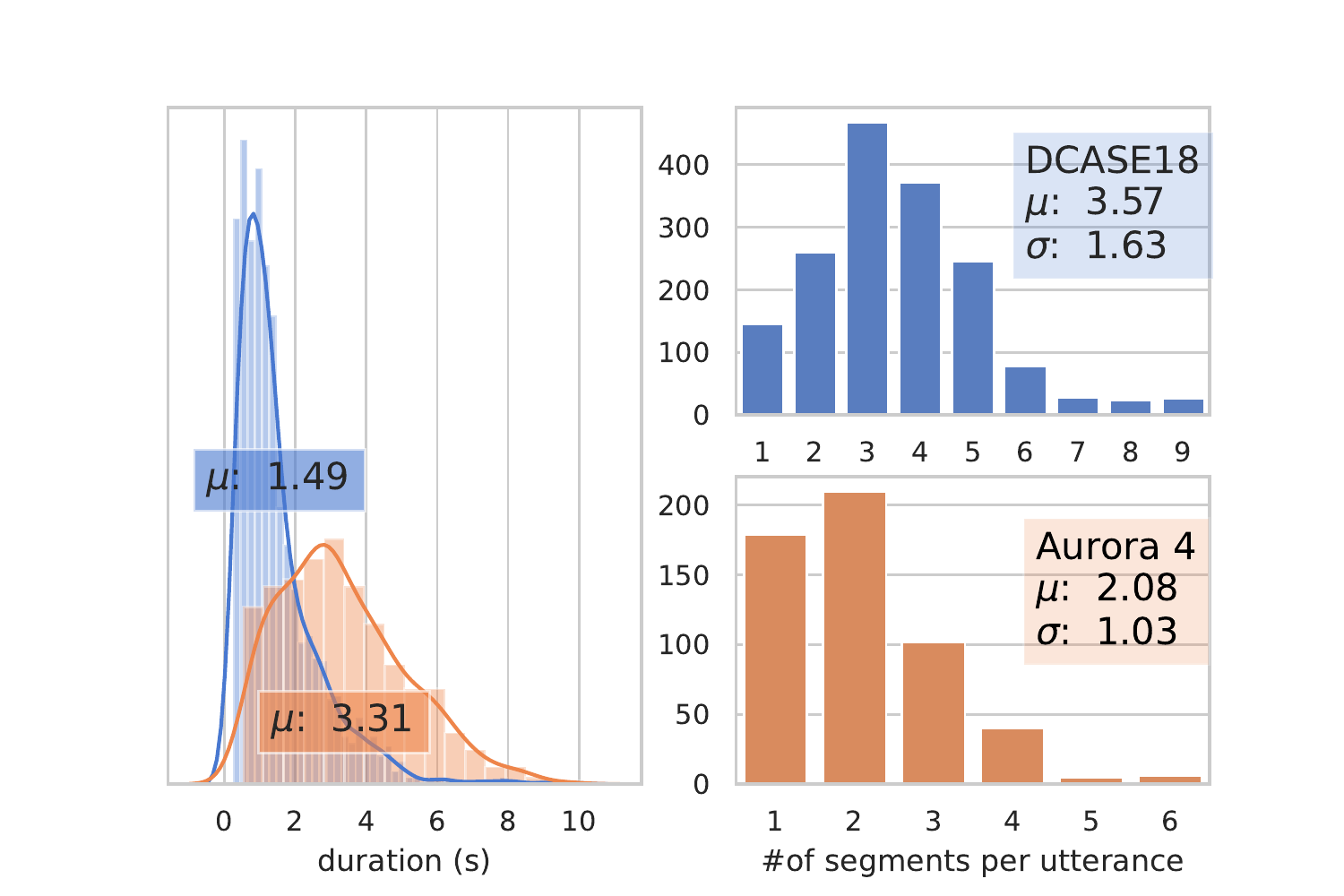}
    \caption{Evaluation data distribution with regards to duration (left) and number of segments per utterance (right), between the Aurora 4 (orange) and DCASE18 (blue) sets. Best viewed in color. }
    \label{fig:dataset_distribution}
\end{figure}

\begin{table*}
\centering
\begin{tabular}{l|l|l||r|r|r|r|r|r}
     \multirow{2}{*}{Testset} & \multirow{2}{*}{Condition} & \multirow{2}{*}{Model} & \multicolumn{5}{c}{Metric}  \\
     & & & F1-macro(\%) & F1-micro(\%) & AUC(\%) & FER(\%) & Event-F1(\%) \\
     \hline\hline 
     \multirow{3}{*}{Aurora 4 (A)} & \multirow{3}{*}{Clean} & VAD-C & \textbf{96.55} & \textbf{97.43} & \textbf{99.78} & \textbf{2.57} & \textbf{78.9} \\
     & & GPV-B & 86.24 & 88.41 & 96.55 & 11.59 & 21.00 \\
     & & GPV-F &  \underline{95.58} &  \underline{95.96} &  \underline{99.07} &  \underline{4.01} &  \underline{73.70} \\
     \hline
     \multirow{3}{*}{Aurora 4 (B)} & \multirow{3}{*}{Syn} & VAD-C & \textbf{85.96} & \textbf{90.28} & \textbf{97.07} & \textbf{9.71} & \textbf{47.5} \\
     & & GPV-B & 73.90 & 75.75 & 89.99 & 24.25 & 8.0 \\
     & & GPV-F &  \underline{81.99} &  \underline{84.26} &  \underline{94.63} &  \underline{15.74} &  \underline{35.4} \\
     \hline
     \multirow{3}{*}{DCASE18} & \multirow{3}{*}{Real} & VAD-C & 77.93 &  \underline{78.08} & 87.87 & 21.92 &  \underline{34.4} \\
     & & GPV-B &  \underline{77.95} & 75.75 &  \underline{89.12} &  \underline{19.65} & 24.3 \\
     & & GPV-F & \textbf{83.50} & \textbf{84.53} &  \textbf{91.80} & \textbf{15.47} & \textbf{44.8} \\
\end{tabular}
\caption{Best achieved results on each respective evaluation condition. Bold marks best result for the respective dataset, while underlined marks second best.}
\label{tab:results}
\end{table*}

As it can be seen in \Cref{fig:dataset_distribution}, the DCASE18 evaluation datasets differ from the Aurora 4 dataset in terms of average duration spoken (1.49 s vs. 3.31 s), as well as number of spoken segments within an utterance (3.87 vs. 2.08).

\subsection{Evaluation Metrics}
\label{ssec:eval}

\paragraph*{Frame-level} For frame-level evaluation, we utilize frame macro/micro averaged F1 scores (F1-macro, F1-micro), Area Under the Curve (AUC)~\cite{ROC_AUC}, and frame error rate (FER). 

\vspace{-10pt}
\paragraph*{Segment-level} For segment-level evaluation we utilize  event-based F1-Score (Event-F1)~\cite{Mesaros2016,Bilen2019}.
Event-F1 calculates whether onset, offset, and the predicted label  overlaps with the ground truth, therefore being a measure for temporal consistency.
We set a t-collar value according to WSSED research~\cite{Serizel2018} to 200 ms to allow an onset prediction tolerance and further permit a duration discrepancy between the reference and prediction of 20\%. 




\subsection{Setup}
\label{ssec:setup}

Regarding feature extraction, all experiments used $64$-dimensional log-Mel power spectrograms (LMS) in this work.
Each LMS sample was extracted by a $2048$ point Fourier transform every $20$ ms with a window size of $40$ ms using a Hann window.
During training, zero padding to the longest sample-length within a batch is applied, whereas, during inference, a batch-size of $1$ is utilized, meaning no padding.

The training criterion for all experiments between the ground truth $\hat{y}$ and prediction $y$ is cross-entropy \Cref{eq:bce} for all samples $N$.
\begin{equation}
    \label{eq:bce}
    \mathcal{L}(\hat{y}, y) = -\frac{1}{N} \sum_{n}^N \hat{y}_{n} \log(y_n) + (1-\hat{y}_n)\log(1-y_{n}) 
\end{equation}
Linear softmax~\cite{Wang2018,dinkel2019duration} (\Cref{eq:linear_softmax}) is utilized as temporal pooling layer that merges frame-level probabilities $y_t(e) \in \left[ 0,1 \right]$ to a single vector representation $y(e) \in \left[ 0,1 \right]^E$.
\begin{equation}\label{eq:linear_softmax}
          y(e) = \frac{\sum_{t}^T y_t(e)^2}{\sum_{t}^T y_t(e)}
\end{equation}
\vspace{-10pt}
\paragraph*{GPVAD} 
The available training data was split into a label-balanced 90\% training and a 10\% held-out set for model training using stratification~\cite{sechidis2011stratification}. 
Due to the inherent label-imbalance within Audioset, sampling is done such that each batch contains evenly distributed clips from each label.
Training uses Adam optimization with a starting learning rate of $1\mathrm{e}{-4}$, a batch size of 64, and terminates after seven epochs if the criterion did not decrease on the held-out dataset.

\vspace{-10pt}
\paragraph*{VAD-C}
VAD-C training utilizes a batch size of 20, whereas the loss (\Cref{eq:bce}) is ignored for padded frames. 
The learning rate is set to $1\mathrm{e}{-5}$, and SGD is used for model optimization. 
Training target labels are obtained by force alignment from a Kaldi trained ASR HMM model~\cite{povey2011kaldi}.

\vspace{-10pt}
\paragraph*{Post-processing}
\label{para:post_processing}

During inference, post-processing is required in order to obtain hard labels from class-wise probability sequences ($y_t(e)$).
We hereby use double threshold~\cite{dinkel2019duration,Kong2018} post-processing, which uses two thresholds $\phi_{\text{low}}=0.1, \phi_{\text{hi}}=0.5$.

\section{Results}
\label{sec:results}



Our results can be seen in \Cref{tab:results}. 
Firstly, we show that our VAD-C model is capable of performing on an equal footing to other deep neural network approaches~\cite{Tong2016}.
Comparing VAD-C with GPV-B/F, it can be seen that VAD-C is indeed the best performing model given our metrics for clean and synthetic noise datasets.
However, evaluation on the real-world dataset reveals a different picture.
Here, VAD-C seems to be struggling against the naive GPV-B approach (AUC 87.87 vs. 89.12, FER 21.92 vs. 19.65), indicating that VAD-C is more likely to misclassify speech in the presence of real-world noise.
Moreover, in real-world scenarios, GPV-F is outperforming VAD-C for each proposed metric.
Our proposed GPV-F approach can also be seen to be consistently noise-robust since its performance difference between synthetic noise and real-world scenarios is minor.
\begin{figure}
    \centering
    \includegraphics[width=\linewidth]{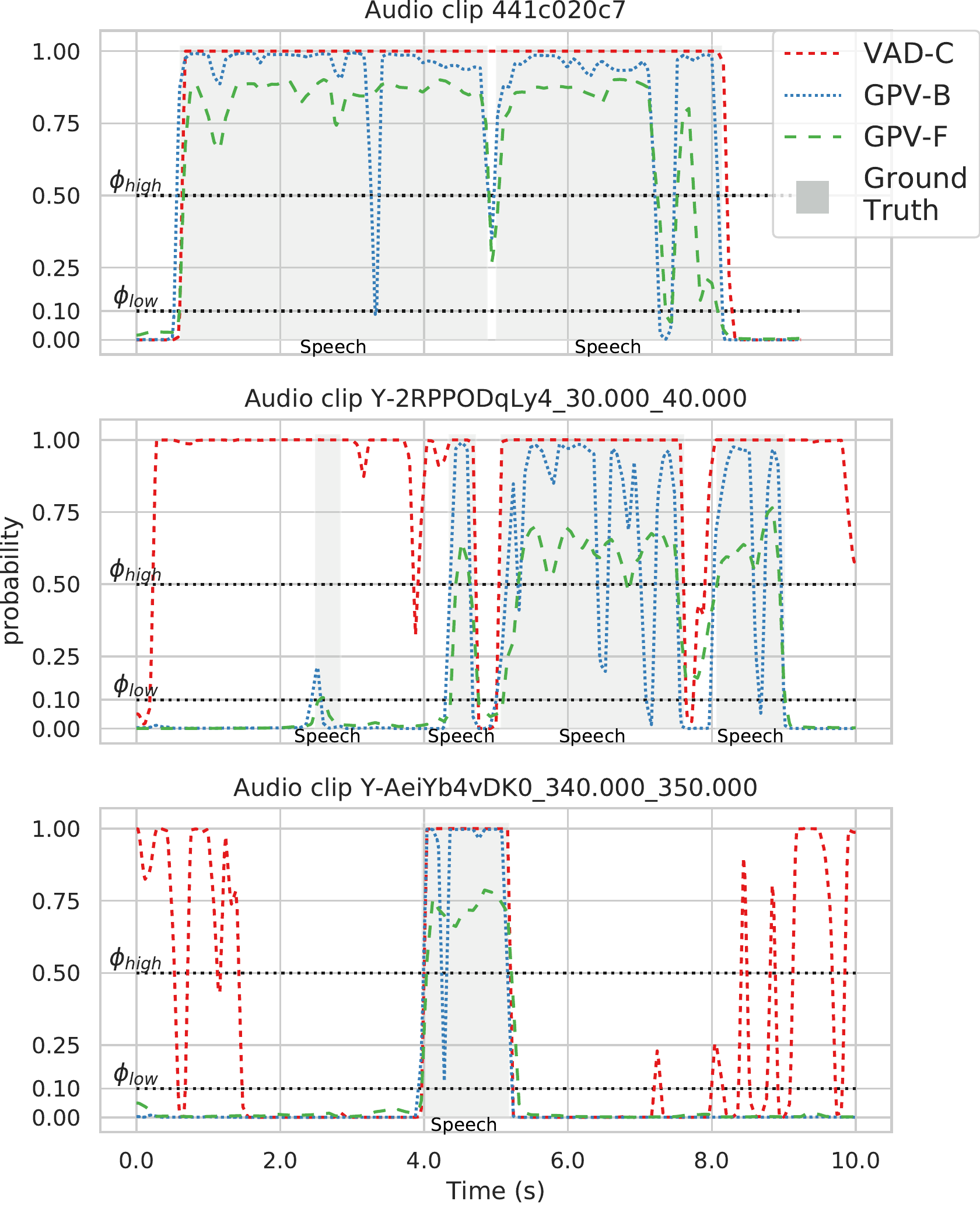}
    \caption{Per-frame probability output for three sample clips, with visualized speech occurrence (boxed, gray). (Top) Contains a clip from Aurora 4 (B); (Center) contains a musician playing a guitar (DCASE18); (Bottom) contains somebody talking with background noises (DCASE18). Post-processing thresholds $\phi_{high},\phi_{low}$ are indicated. Best viewed in color. }
    \label{fig:prob_output}
\end{figure}

Even though GPV-B is, on average underperforming against the other two approaches, one should note that it is the least costly system, since labeling data for GPV-B is essentially a binary question whether one heard any speech within a clip, making this approach capable of cheaply scaling to large data.
We conclude that the GPVAD models trained with only clip-level labels are capable of competing trained on frame-level labels.

\paragraph*{Quantitative Results}
In order to visualize model-specific behavior, three clips (one Aurora 4 Noisy, two DCASE18) were sampled from the testing set, and per-frame output probabilities are shown for each model seen in \Cref{fig:prob_output}.
In the case of the synthetic Aurora 4 test at the top, we can see that our GPVAD models are capable of modeling short pauses between two speech segments, at which VAD-C fails, yet both GPVAD models could not correctly estimate the second speech segments end.
The center sample further demonstrated a typical VAD-C problem in real-world scenarios: it is unable to distinguish between foreground events (here Guitar) and active speech for a majority of the utterance.
Especially the bottom sample exemplifies this problem: VAD-C starts to predict speech, where there is none, while both GPVAD models are capable of distinguishing any background noises from speech.
Please note that the bottom clip contains laughter at the end, which VAD-C classifies as speech.
In our future work, we would like to further extend the scope of GPVAD training by utilizing larger training data (e.g., unbalanced AudioSet).


\section{Conclusion}
\label{sec:conclusion}

This paper introduces a noise-robust VAD approach by utilizing weakly labeled sound event detection.
Two GPVAD systems are investigated: GPV-B, trained on binary speech and non-speech pairs only, as well as GPV-F, which utilizes all 527 AudioSet labels.
Our evaluation protocol thoroughly compares our proposed GPVAD approach to traditional VAD utilizing five distinct metrics.
Results indicate that GPV-B, even though trained on clip-wise, unconstrained speech, can be used to detect spoken language, without requiring clean, frame-labeled training data.
Further, while GPV-B/F both fall short in clean and synthetic noise scenarios against VAD-C, they excel at stable predictions for real-world data.
Specifically it can be seen that our proposed approach is robust in its performance across the synthetic and real-world noise datasets.
Our best performing model, GPV-F outperforms traditional supervised VAD approaches by a significant margin on real-world data, culminating in an absolute increase of 5.57\% F1-macro, 6.45\% F1-micro, 3.93\% AUC, 6.45\% FER and 10.4\% Event-F1.

\section{Acknowledgements}

This work has been supported by National Natural Science Foundation of China (No.61901265) and Shanghai Pujiang Program (No.19PJ1406300). Experiments have been carried out on the PI supercomputer at Shanghai Jiao Tong University.

\bibliographystyle{IEEEtran}

\bibliography{ref}

\end{document}